\documentclass{nature}
\usepackage{graphicx}
\makeatletter
\let\saved@includegraphics\includegraphics
\AtBeginDocument{\let\includegraphics\saved@includegraphics}
\renewenvironment*{figure}{\@float{figure}}{\end@float}
\makeatother

\usepackage{multirow}
\usepackage{subfigure}
\usepackage{mathtools}
\usepackage[colorlinks=true,linkcolor=blue,citecolor=black,urlcolor=blue]{hyperref}
\usepackage{authblk}
\usepackage[labelfont=bf]{caption}
\usepackage[figurename=Figure]{caption}
\usepackage{siunitx}
\usepackage[dvipsnames]{xcolor}
\usepackage{soul}
\usepackage{xfrac}
\usepackage{amsmath}
\usepackage{soul,xcolor}
\usepackage{lineno}

\begin{document}

\setstcolor{red}

\title{Universal patterns of skyrmion  magnetizations unveiled by defect implantation}
\author[1]{Imara Lima Fernandes}
\affil{Peter Gr\"{u}nberg Institut and Institute for Advanced Simulation, Forschungszentrum J\"{u}lich and JARA, D-52425 J\"{u}lich, Germany}
\author[1,2*]{Samir Lounis}
\affil{Faculty of Physics, University of Duisburg-Essen and CENIDE, 47053 Duisburg, Germany}
\affil[*]{s.lounis@fz-juelich.de}

\maketitle


\begin{abstract}
Skyrmions are spin-swirling textures hosting wonderful properties with potential implications in information technology. Such magnetic particles carry a magnetization, whose amplitude is  crucial to establish them as robust magnetic bits, while their topological nature gives rise to a plethora of exquisite features such as topological protection, the skyrmion and topological Hall effects as well as the topological orbital moment. These effects are all induced by an emergent magnetic field directly proportional to the three-spin scalar chirality, $\chi= (\mathbf{S}_i\times\mathbf{S}_j)\cdot \mathbf{S}_k$, and shaped by the peculiar spatial dependence of the magnetization. Here, we demonstrate the existence of novel chiral magnetizations emerging from the interplay of spin-orbit interaction and either $\chi$ or the two-spin vector chirality $\boldsymbol{\kappa} = \mathbf{S}_i\times\mathbf{S}_j$. By scrutinizing correlations among the spin, orbital (trivial and chiral) magnetizations, we unveil from ab-initio universal patterns, quantify the rich set of magnetizations carried by single skyrmions generated in PdFe bilayer on Ir(111) surface and demonstrate the ability to engineer their magnitude via controlled implantation of impurities. We anticipate that our findings can guide the design of disruptive storage devices based on skyrmionic bits by encoding the desired magnetization with  strategic seeding of defects.
\end{abstract}


\section*{Introduction}

The identification of universal patterns characterizing properties of materials is crucial for various fields of science. This enables to design materials with the right combination of elements to achieve desired characteristics. A fundamental example is the Mendeleev periodic table that consists of a graphical representation of chemical elements, which depending on their atomic numbers and location in the table share trends with periodic patterns, such as 
electron negativity, affinity, metallicity and cohesion.  
In the context of magnetism, Hund's rules provide generic dependencies of the spin and orbital moments as function of the filling of electronic states. 
The spin moment follows an inverse parabolic behavior with a maximum expected at half-filling of the $d$-states, while the orbital moment exhibits an S-shape passing through zero at half-filling with early (late) transition elements having an orbital moment antiferromagnetically (ferromagnetically) aligned to the spin moment. The Slater-Pauling rule~\cite{Slater36,Pauling38} illustrates another primary general prescription in magnetism, which uses  the electronic filling as a fundamental descriptor dictating how the total magnetization of an alloy is shaped by the amount of valence electrons injected to or removed   upon alteration of the concentration of the atomic building-blocks. Instead of the electronic filling, the radius of the 3$d$ electron shell acts as the descriptor defining the celebrated Bethe-Slater curve~\cite{Slater30,Slater30b,BetheSommerfeld1933}, a common textbook example, which is fundamental to understand the magnetic properties and magnetic ordering of transition metal materials.

\begin{figure}[h!]
\centering
    \includegraphics[width=1.\linewidth]{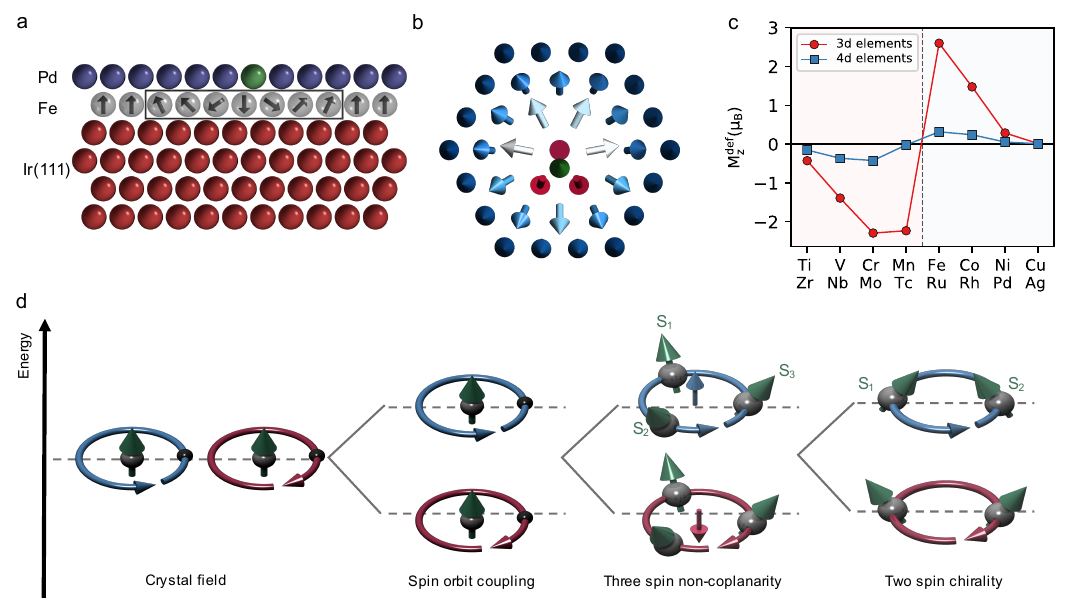}
    \caption{\textbf{Various orbital magnetizations emerging in a magnetic skyrmion.} \textbf{a}  Schematic drawing of the cross-section of PdFe/Ir(111) surface, which incorporates an atomic defect (green sphere) implanted in the Pd overlayer. \textbf{b} Spin structure of a magnetic skyrmion interacting with a Ti defect. The green arrow and sphere represent the defect. \textbf{c} Spin moments of the series of 3$d$ and 4$d$ atomic defects when interacting with the ferromagnetic PdFe bilayer. Positive (negative) values indicate a ferromagnetic (antiferromagnetic) alignement of the impurity spin moment with respect to the substrate magnetization. \textbf{d} Schematic representation of various mechanisms affecting the electronic level splitting from which emerge distinct contributions to the orbital moment. In a solid, the crystal field quenches the orbital moment, while spin-orbit coupling lifts the orbital degeneracy. Even without spin-orbit coupling, an orbital moment (3-spin chiral orbital moment) emerges from the non-coplanarity of three magnetic moments. With spin-orbit coupling an additional chiral orbital moment is enabled by the two-spin vector chirality (2-spin chiral orbital moment). A third chiral orbital moment (not shown) can be induced by a plaquette of four spins, which can be interpreted as an interference term depending on spin-orbit interaction and the three-spin scalar chirality. }
    \label{fig:Figure1}
\end{figure}

In this work, we unveil a universal behavior characterizing the magnetization carried by complex spin-textures. We exemplify our findings by addressing magnetic skyrmions~\cite{Bogdanov1989,Roessler2006}, which are topological objects confined in two-dimension~\cite{Nagaosa2013} and heavily prospected as a mean of encoding magnetic bits for information technology~\cite{Fert2013,Fert2017}. Here the magnetic bit is defined by the magnetization carried by the skyrmion, which is different from the surrounding background. Clearly the ability to engineer the amplitude of the magnetization is paramount for potential device applications. Moreover, it is the peculiar shape of the magnetization that defines the emergent magnetic field giving rise to the topological properties characterizing skyrmions, such as the topological charge as well as the topological and skyrmion Hall effects~\cite{Nagaosa2013}. 
 Currently three types of magnetization are discussed in the literature (see Fig.~\ref{fig:Figure1}d). Besides the spin and usual orbital magnetization (UOM) induced respectively by the exchange splitting and spin-orbit interaction (SOI), which lift spin and orbital degeneracy, the chiral orbital magnetization (COM) was recently introduced as a quantity emerging independently from SOI due to magnetic non-coplanarity~\cite{Shindou01,dosSantosDias2016,Hanke2016,dosSantosDias2017,Bouaziz2018,Lux2018,Brinker2020b}. Non-vanishing ground-state currents~\cite{Tatara03}, giving rise to an orbital magnetization, develop  proportionally to the three-spin scalar chirality $\chi_{ijk} = (\mathbf{S}_i\times\mathbf{S}_j)\cdot\mathbf{S}_k$, i.e. the solid angle defined by a plaquette of  three magnetic moments with $\mathbf{S}_i$ being the unit vector of the magnetic moment carried by atom $i$. $\chi$ induces the emergent magnetic field~\cite{Bruno2004,Everschor14} responsible for various Hall effects~\cite{Bruno2004,Lux2020,Bouaziz2021}, triggers electromagnetic induction~\cite{Nagaosa2019,Yokouchi2020} and is distinct from the two-spin chirality vector $\boldsymbol{{\kappa}}_{ij} = \mathbf{S}_i\times\mathbf{S}_j$ associated with the Dzyaloshinskii-Moriya interaction~\cite{Dzyaloshinsky1958,Moriya1960}. Ultimately, the total three-spin scalar chirality summed up over the whole  topological spin-texture with weakly rotating spin moments is quantized by being proportional to the topological charge, which enforces a topological behavior on key transport phenomena and promotes the chiral orbital magnetization to a topological one~\cite{dosSantosDias2016}. We refer to the aforementioned chiral orbital magnetization as the three-spin chiral orbital magnetization (3-spin COM).
 
Utilizing fundamental concepts in multiple-scattering theory, we demonstrate the emergence of new chiral orbital magnetizations triggered by SOI and non-collinearity (Fig.~\ref{fig:Figure1}d) via mechanisms involving multi-spin plaquettes. The first one is directly proportional to the two-spin chirality $\boldsymbol{\kappa}$, called then 2-spin COM, and therefore changes sign upon chirality switching, also identified semi-classically~\cite{Lux2018}. The second one looks like an interference effect produced by both the three-spin chirality $\chi$ and SOI and emerges from a four-spin plaquette. It is therefore refered to as the 4-spin COM. We quantify from first-principles the various magnetizations carried by a realistic few-nanometers wide N\'eel-type skyrmion known to emerge in fcc-PdFe bilayer deposited on Ir(111) surface~\cite{Romming2013,LimaFernandes2018,Fernandes2020,Fernandes2020a,Fernandes22} and explore the magnetization responses to single atomic defects implanted in the Pd overlayer (see Figure~\ref{fig:Figure1}a and Methods). We demonstrate that the impurities can inject large magnetizations into the skyrmion. As defects we consider  3$d$ (Ti, V, Cr, Mn, Fe, Co, Ni) that can carry significant spin moments (Figure~\ref{fig:Figure1}c) and 4$d$ (Zr, Nb, Mo, Tc, Ru, Rh) transition metal atoms with much weaker spin moments as well as Cu and Ag atoms, which are positioned close to the core of the skyrmion (Figure~\ref{fig:Figure1}a-b). We identify common patterns in the spin and orbital magnetization as function of the atomic number of the defects, i.e. the impurities electronic filling, which are surprisingly also  executed by the SO-independent chiral orbital magnetization. The unveiled patterns  are however distinct for 3$d$ and 4$d$ impurities and find their origin in the magnetic interactions dictating the deformation of the skyrmionic texture. 
 We propose the three-spin scalar chirality as a common descriptor for the spin, orbital and chiral magnetizations, which can be probed by the imposed impurities.  While the conventional SO-induced orbital magnetization is expected to be linear with the spin magnetization, the three-spin scalar chirality is a cubic function of the magnetization, which explains the unveiled dependencies and defines the generic trends.

\section*{Results}

\subsection{Chirality-induced magnetizations.}

The shape of a magnetic skyrmion is modified as soon as a defect is implanted at its vicinity. This naturally modifies the total magnetic moment carried by the non-trivial spin-texture. As demonstrated in Supplementary Note 1 and illustrated in Supplementary Figure 1, we expect new orbital moments induced by the non-coplanarity and/or chirality  of a spin-texture besides the three types of magnetic moments discussed in the literature: (i) the spin, (ii) the orbital and (iii) chiral (topological) orbital magnetic moments.  Utilizing multiple-scattering theory, we demonstrate that the orbital moment $\mathbf{m}^\text{orb}$ at a given atomic site $i$ can be approximately decomposed into four contributions categorized in terms of multi-spin inducing processes with and without SOI:

\begin{equation}
    \mathbf{m}^\mathrm{orb} =  \mathbf{m}^\mathrm{orb}_\mathrm{1-spin} + \mathbf{m}^\mathrm{orb}_\mathrm{2-spin} +\mathbf{m}^\mathrm{orb}_\mathrm{3-spin}+  \mathbf{m}^\mathrm{orb}_\mathrm{4-spin}.
\end{equation}

We start with the most basic contribution, $\mathbf{m}^\mathrm{orb}_\mathrm{1-spin}$, which is the usual orbital moment (UOM) induced by SOI and is collinear with the atomic spin moment. It is thus an orbital moment that primarily arises from a single spin term, $\mathbf{S}_i$, in contrast to the two-spin induced orbital moment (2-spin COM), $\mathbf{m}^\mathrm{orb}_\mathrm{2-spin}$, which is chiral since it is proportional to the two-spin chirality $\boldsymbol{\kappa}$, i.e. $\propto \left(\mathbf{S}_i \times \mathbf{S}_j\right)$. Owing to the chiral nature of this term, rotating the moments clockwise or counter-clockwise modifies the orbital moment in an opposite fashion. It is linear with SOI and is finite when inversion symmetry is broken. The three-spin induced $\mathbf{m}^\mathrm{orb}_\mathrm{3-spin}  \propto \chi$ is the chiral orbital moment (3-spin COM) triggered by the three-spin scalar chirality $\chi$ ($\propto (\mathbf{S}_i \cdot \left(\mathbf{S}_j \times \mathbf{S}_k\right)$), while the four-spin induced orbital moment (4-spin COM), $\mathbf{m}^\mathrm{orb}_\mathrm{4-spin}$, is a sort of interference  moment enabled by SOI and proportional to the three-spin scalar chirality ($\mathbf{S}_i \chi$). Interestingly, among the four contributing terms, only $\mathbf{m}^\mathrm{orb}_\mathrm{3-spin}$ does not request SOI to be finite. We note that a five-spin chirality term can induce an orbital moment~\cite{Bouaziz2018} even without SOI, but it is found negligible in the current study. Furthermore, for a magnetically collinear state, all chiral orbital magnetizations cancel out.

\subsection{Unveiling skyrmion's magnetization patterns by defect implantation.}

\begin{figure}[h!]
\centering
    \includegraphics[width=1.\linewidth]{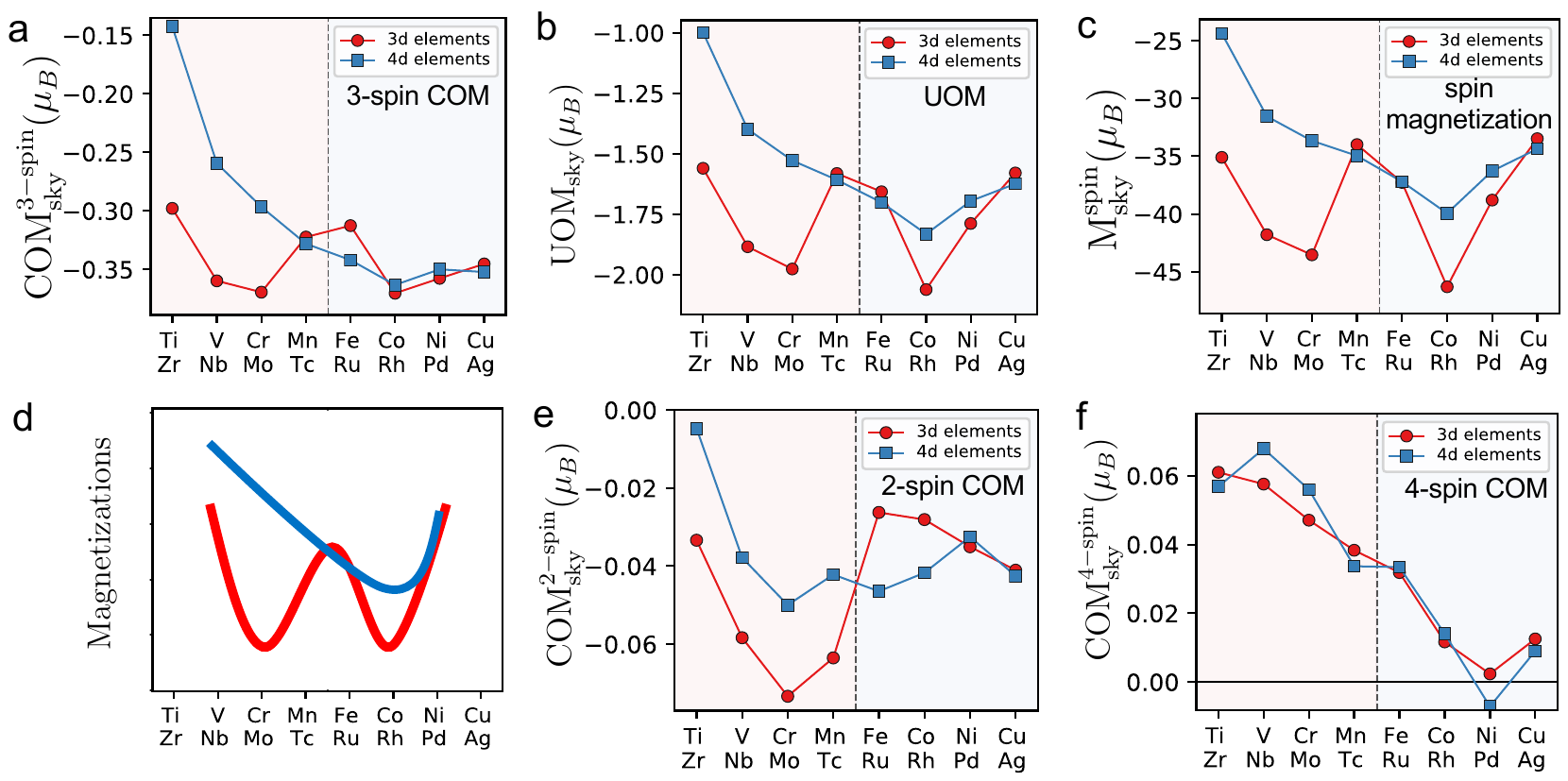}
    \caption{\textbf{Universal trends of distinct skyrmion-carried magnetizations unveiled by implanted atomic defects.} \textbf{a} Total 3-spin chiral orbital magnetization (COM), \textbf{b} usual orbital magnetization (UOM), \textbf{c} spin magnetization,  \textbf{e} 2-spin COM and \textbf{f} 4-spin COM. \textbf{d} Schematic illustration of the magnetizations, chiral or not,  pertaining to a single skyrmion as function of the electronic filling across the transition metal series.}
    \label{fig:Figure2}
\end{figure}

Following the method of extraction described in Supplementary Note 1, each of these orbital moments together with the spin moment summed up over a whole magnetic skyrmion are presented in Fig.~\ref{fig:Figure2}, which gives an insight on the sensitivity of the total skyrmion's magnetization to the electronic filling of atomic defects implanted at the vicinity of the skyrmmion's core. We note that the skyrmion magnetization is defined as the difference in the total magnetic moment with and without the skyrmion, which is here of  N\'eel type and has a diameter of about \qty{2.2}{n\meter}. Overall the net skyrmion magnetization points along the z-direction. 

The 3-spin COM  carried by the whole skymion can be dramatically modified by the presence of atomic defects (see Fig.~\ref{fig:Figure2}a). Co or Cr impurities can enhance it by about 6\% while a strong reduction can be achieved by Ti. Intriguingly, one can clearly see that on the one hand 3$d$ atomic defects induce 3-spin COMs exhibiting a W-shape as function of the impurities atomic number. The curve shows two minima located at Cr and Co separated by a maximum found at Fe.  On the other hand, 4$d$ defects trigger 3-spin COMs following a V-shape, with a single minimum at Rh. Whilst the late transition elements (say starting from Tc) induce moments similar to those of the 3$d$ defects, early elements (Zr, Nb and Mo) strongly reduce the skyrmion's chiral orbital moment.

Surprisingly the identified trends, W- and V-shapes for respectively 3$d$ and 4$d$ impurities as schematically illustrated in Fig.~\ref{fig:Figure2}d, are exactly followed by the UOM conventionally induced by SOI and the spin magnetization of the skyrmion (Fig.~\ref{fig:Figure2}b-c). These results indicate some general pattern behavior connecting the three types of magnetizations and a common hidden link to the three-spin scalar chirality. In contrast to the 3-spin COM, the UOM is much larger and can change by almost 1$\mu_B$ across the different impurities while the spin magnetization exhibits alterations reaching about 25$\mu_B$. This highlights the potential of using atomic defects to engineer the magnetization carried by a single skyrmion. The newly proposed COM, enabled by SO and  the two-spin chirality (Fig.~\ref{fig:Figure2}e) or the three-spin chirality (Fig.~\ref{fig:Figure2}f), are one order of magnitude smaller than the SO-independent chiral orbital moment. The shapes of the 2-spin and 4-spin COM for the 4$d$ impurities seem similar to that obtained for the rest of the magnetization, with a minimum observed for Mo (2-spin COM) and Pd (4-spin COM). The 2-spin COM induced by the 3$d$ atomic defects exhibits the W-shape, with a minimum obtained for Cr, while the second minimum has been pushed to the limit of the series of transition elements (Cu). For the 4-spin COM, the first minimum induced by  the early 3$d$ transition elements in the rest of magnetizations disappears leading to a V-like shape with a single minimum for Ni. As aforementioned, the 4-spin COM is a sort of interference term, which is in the particularly investigated skyrmion weak in comparison to the other studied orbital magnetizations. In the following, we focus our discussion on the spin magnetization, the UOM and 3-spin COM, which are significantly larger in magnitude than the 2-spin and 4-spin COM. The two-dimensional maps of the spin and orbital moments across the skyrmion with and without impurities are illustrated in Supplementary Figures 3-8.

\subsection{Origin of the spin-magnetization patterns.}
\begin{figure}[h!]
\centering
    \includegraphics[width=1.\linewidth]{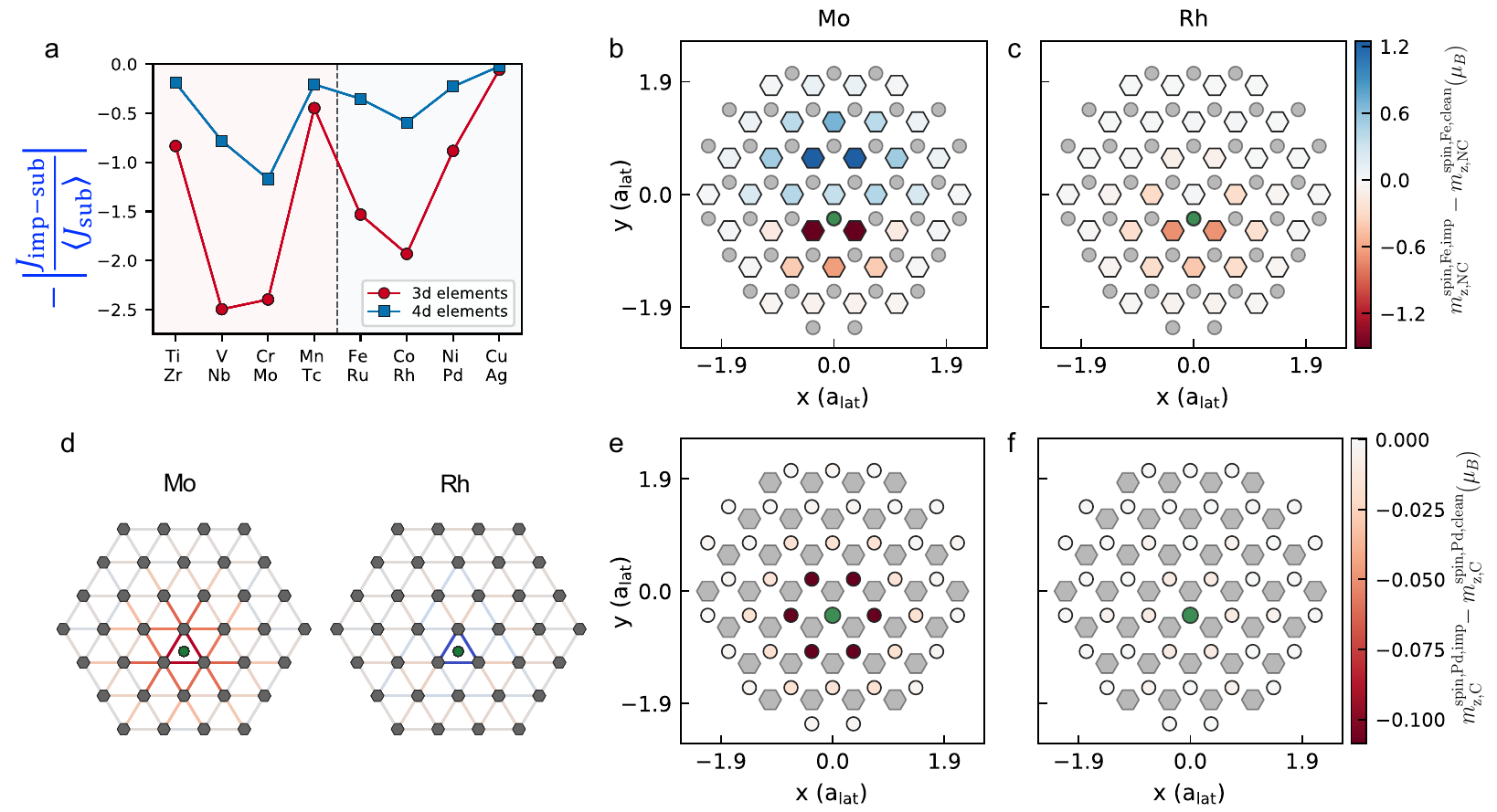}
    \caption{\textbf{Impact of substrate-defect interactions on the trends of skyrmion's magnetization.} \textbf{a} Ratio between the next neighboring impurity-substrate and Fe-Fe magnetic exchange interactions for the case of 3$d$ and 4$d$ atomic defects as a function of the impurities atomic number. Change of the substrate Fe magnetic moments due to the presence of a \textbf{b} Mo and \textbf{c} Rh defects when comparing with a defect-free skyrmion. \textbf{d} Local change of the first nearest neighbour exchange interaction within the Fe-layer due to Mo- and Rh-defects. The red (blue) color of the bond between the atoms indicates the exchange interaction decreased (increased) as compared to the defect-free substrate. Modification of the  overlayer Pd atoms induced by \textbf{e} Mo and \textbf{f} Rh defects as obtained in the collinear state.}
    \label{fig:Fig3}
\end{figure}

To understand the origin of the magnetization patterns and the dissimilarity between those pertaining to 3$d$ versus 4$d$ impurities, we explore how the latter impact the spin magnetization of the underlying Fe layer, which hosts the non-trivial spin-textures. Two mechanisms are at play: (i) the competition between the defect-substrate magnetic interaction with the intralayer Fe-Fe magnetic interaction, and (ii) the alteration of the latter by the presence of impurities. In contrast to 4$d$ impurities, 3$d$ atomic defects carry a large spin magnetic moment (see Fig.~\ref{fig:Figure1}c) and impose a strong magnetic exchange interaction to the substrate. As shown in Supplementary Note 2 for a simple model, the z-magnetization is expected to be proportional to the ratio between the impurity-substrate magnetic interaction and the nearest neighboring interaction within the Fe layer, $|\mathrm{J}_\mathrm{Fe-imp}/\mathrm{J}_\mathrm{Fe-Fe}|$~\cite{Lounis2007}. This is calculated for the investigated skyrmion and plotted in Fig.~\ref{fig:Fig3}a, which reveals the aforementioned W-shape. On the same figure, we deliberately show what the Heisenberg model predicts for the 4$d$ impurities, although the model is not appropriate since the magnetic moments are much smaller than the 3$d$ ones and are more sensitive to both the  surrounding and spin-rotation. The latter ingredients prevent the use of the Heisenberg model, which interestingly predicts a behavior similar to that induced by the 3$d$ atomic defects.

The origin of the  V-shape found for the 4$d$ impurities lies in the way the effective magnetic interaction among the substrate Fe atoms is affected by the presence of the defects. In the defect-free case, the overlayer Pd atoms enhance the nearest neighboring Fe-Fe magnetic interaction in a linear fashion~\cite{LimaFernandes2018}. Indeed the Pd atoms are characterized by a ferromagnetic Stoner susceptibility, which at the vicinity of the magnetic substrate carry an induced moment facilitated by the Stoner criterion for magnetism. The bare nearest neighboring Fe-Fe magnetic interaction is then enhanced by 16\% in average~\cite{LimaFernandes2018}. A similar enhancement is expected for impurities that tend to couple ferromagnetically to the substrate, which is the case for example of Rh (see Fig.~\ref{fig:Fig3}d). For instance, when replacing the Pd atom by the Rh, the magnetic moment locally increases from $\qty{0.3}\mu_\text{B}$ to $\qty{0.9}\mu_\text{B}$. This stiffens the skyrmion and enlarges its core as it can be grasped from Fig.~\ref{fig:Fig3}c, which leads to a negative contribution to the z-component of the magnetization and reduces it. Mo impurity, however, couples antiferromagnetically to the Fe substrate, this affects the surrounding Pd atoms by reducing their magnetization as shown in  Fig.~\ref{fig:Fig3}e, which can be compared to the Rh-case in Fig.~\ref{fig:Fig3}f. This diminishes the magnitude of the effective Fe-Fe magnetic interaction in comparison to the defect-free scenario as illustrated in Fig.~\ref{fig:Fig3}d. In this case, the outer-region of the skyrmion, contributing positively to the z-magnetization, wins even more over the core region (Fig.~\ref{fig:Fig3}b) and the total magnetization is enhanced. For completeness, two-dimensional maps of the spin moments of Pd overlayer atoms are presented in Supplementary Figures 9-11.

\subsection{Correlation between spin, orbital and chiral magnetizations of skyrmions.} 
\begin{figure}[h!]
\centering
    \includegraphics[width=1.\linewidth]{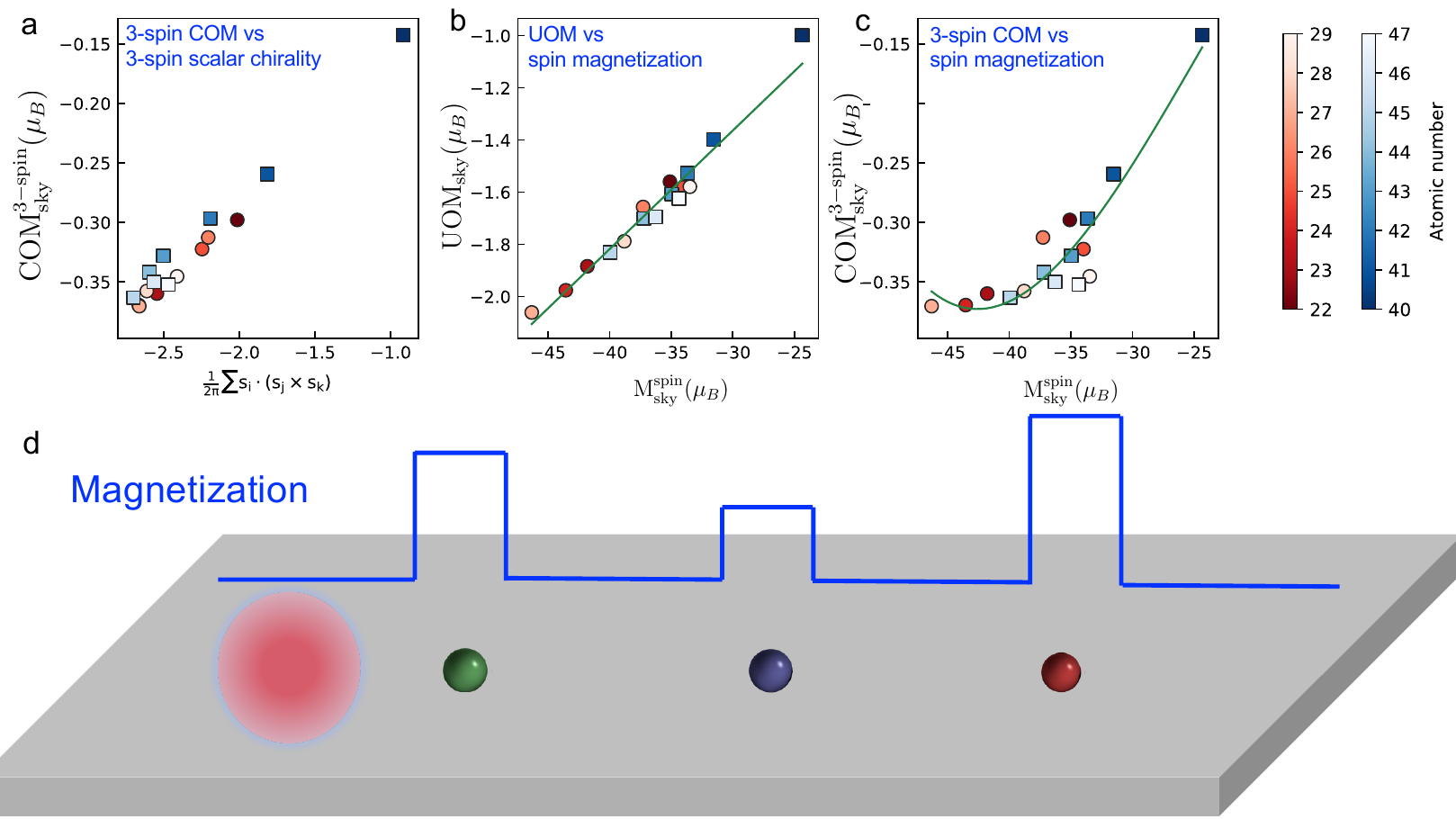}
    \caption{\textbf{Universal correlations between skyrmion's magnetizations and device concept.} \textbf{a} The skyrmion's 3-spin COM as function of the total 3-spin scalar chirality collected from every magnetic profile induced by defects of the 3$d$ and 4$d$ series.  \textbf{b} The skyrmion's UOM and \textbf{c} 3-spin COM as function of the skyrmion's spin magnetization. While  the 3-spin COM and UOM exhibit a linear behavior, as expected, with respectively the 3-spin scalar chirality and spin magnetization, a cubic function describes the relation between  the 3-spin COM and spin magnetization. \textbf{d} Based on the universal dependence of the skyrmion's magnetization with respect to the atomic number of an impurity, a device is envisaged with a controlled implantation of defects to engineer the amplitude of the magnetization. The correlation among the different types of skyrmion's magnetizations enables to monitor,  design and quantify the chiral orbital contributions once the spin magnetization is measured. 
    }
    \label{fig:Figure4}
\end{figure}

The previous subsection deals with the origin of the patterns characterizing the spin magnetization carried by a skyrmion. Here, we explain how the unveiled trends are also obeyed by both the UOM and 3-spin COM. To get first insight, we present in Fig.~\ref{fig:Figure4} various correlation plots systematically collected for the different defects implanted close to the core of the magnetic skyrmion. For instance,  Fig.~\ref{fig:Figure4}a shows that the 3-spin COM carried by the skyrmion is linear with the total three-spin scalar chirality, while Fig.~\ref{fig:Figure4}b recovers the expected linear dependence between the UOM and the spin magnetization. An intriguing cubic dependence is, however, found between the three-spin chiral orbital and spin magnetizations (Fig.~\ref{fig:Figure4}c), which can be fitted by the following equation
\begin{equation}
    \text{3-spin COM} = \mathrm{A} \,\mathrm{M}_\text{sky}^\text{spin} + \mathrm{B} \,{\mathrm{M}_\text{sky}^\text{spin}}^2+ \mathrm{C} \,{\mathrm{M}_\text{sky}^\text{spin}}^3, \label{eq:universal}
\end{equation}
where $\mathrm{A} = -1.64 10^{-2}$, $\mathrm{B} = -1.39 10^{-3}\mu_B^{-1}$  and $\mathrm{C} = 1.87 10^{-5} \mu_B^{-2}$ are material-dependent parameters. We show analytically in Supplementary Note 3 that $\mathrm{A} = \frac{\sqrt{3}}{2}$, $\mathrm{B} = 0$ and $\mathrm{C} = - \frac{\sqrt{3}}{18}$ for a non-collinear trimer in a N\'eel state having initially its three magnetic moments lying in-plane, which simultaneously interact with an atomic magnetic defect. A more involved analytical evaluation comfort the cubic dependence presented in Eq.~\ref{eq:universal} and 
conjecture therefore that it settles a universal correlation between the total 3-spin COM and the spin magnetization as well as the UOM.

While the trends dictating the spin-magnetization are induced by the modification of  magnetic interactions  across the transition atomic defect via electronic filling of the impurities, it is the SOI and the cubic dependence of the three-spin scalar chirality that enforce the same trends on respectively the conventional orbital and chiral magnetization.

\newpage

\section*{Discussion}

In this study, we delve into the intricate realm of the magnetization carried by skyrmions. Hinging on multiple-scattering mechanisms, we unveil  chiral orbital magnetizations,  sparked by multi-spin plaquettes intertwined with  spin-orbit interaction, non-collinearity, two-spin and three-spin chiralities. Employing rigorous first-principles quantification, we explore the magnetic landscape of a realistic Néel-type skyrmion residing in an fcc-PdFe bilayer atop an Ir(111) surface. Our inquiry extends to the realm of single atomic defects in the Pd overlayer, revealing the remarkable ability of impurities, to infuse substantial magnetizations into the skyrmion.

We discern compelling patterns universally followed by the spin and orbital magnetizations (usual and chiral) intricately tied to the atomic characteristics of defects and electronic filling. These patterns  showcase nuances between 3$d$ and 4$d$ impurities, stemming from magnetic interactions that sculpt the deformation of the skyrmionic texture. Dissecting the correlations between the different skyrmion's magnetizations influenced by the implanted defects reveals a cubic relation between the spin and chiral orbital magnetization instead of the linear dependence characterizing the spin and orbital magnetizations. 

By elucidating the universal dependencies governing skyrmion magnetization, we contribute to the foundational principles that govern magnetism, paving the way for establishing overarching trends in the intricate domain of chiral magnetism. Our work anticipates the prospect of quantifying chiral magnetization, an aspect yet unmeasured, by precisely determining spin magnetization, strategically leveraging defects that amplify its magnitude.

We envision the transformative potential of strategically implanting defects into surfaces to engineer and manipulate the magnetization carried by skyrmions. Notably, our findings indicate that 3$d$ impurities yield to a more important amplifying effect of the skyrmion's magnetization compared to their 4$d$ counterparts, presenting a promising avenue for advancing skyrmionic bits in the development of disruptive storage devices.

\begin{methods}

\subsection{Computational details.} 
The simulations are based on density functional theory within the local spin density approximation as implemented in the all-electron full-potential  scalar-relativistic Korringa-Kohn-Rostoker (KKR) Green function method with spin-orbit interaction added self-consistently.~\cite{Bauer2013,Crum2015} The method is ideal to treat within an embedding scheme a single confined magnetic skyrmion interacting with atomic defects all embedded on a perfect magnetic substrate. The skyrmion is hosted by the Fe layer and lives within 37 Fe atoms. Nearest neighbors are included, which lead to an embedding cluster of 124 atoms. The spin-texture characterizing the skyrmion is obtained in a self-consistent fashion. The Pd/Fe/Ir(111) slab consists of a fcc-stacked PdFe bilayer deposited on 34 layers of Ir with atomic positions obtained from ab initio. We assume an angular momentum cutoff at lmax=3 for the orbital expansion of the Green function and the energy contour contained 42 grid points in the upper complex plane with seven Matsubara poles and a Brilloun zone mesh of 30$\times$30 k-points for the self-consistent description of skyrmion-free substrate properties (see more technical details in Refs.~\cite{LimaFernandes2018,Fernandes2020,Arjana2020,Bouhassoune2021,Aldarawsheh22}).

\subsection{Analytical derivation of the new chiral magnetizations.} The concept used and presented in Supplementary Note 1 to establish the existence of orbital magnetizations emerging from multiple-scattering events bears some similarities to that utilized to unveil new chiral multi-site magnetic interactions~\cite{Brinker2019,Lounis2020,Grytsiuk2020,Brinker2020}, hyperfine interaction~\cite{Shehada22}, and magnetoresistances~\cite{Fernandes22}.

\subsection{Data availability} The data that support the findings of this study are present in the paper and the Supplementary Information.

\subsection{Code availability} The KKR code is a rather complex ab-initio DFT-based code, which is in general impossible to use without proper training on the theory behind it and on the practical utilization of the code. We are happy to provide the latter code upon request.

\end{methods}

\begin{addendum}

\item We thank Moritz Wineterott for fruitful discussions. This work is supported by the Deutsche For\-schungs\-gemeinschaft (DFG) through SPP 2137 ``Skyrmionics'' (Project LO 1659/8-1). We acknowledge the computing time granted by the JARA-HPC Vergabegremium and VSR commission on the supercomputers JURECA at Forschungszentrum Jülich and at the supercomputing centre of RWTH Aachen University.

\item[Author contributions] 
S.L. initiated, designed, and supervised the project. I.L.F. performed the first-principles calculations and post-processed the data. S.L. provided the analytical derivations of the various orbital magnetizations. I.L.F. and S.L. discussed the results and S.L. wrote the manuscript.

\item[Competing Interests] The authors declare no competing interests.

\item[Correspondence] Correspondence and requests for materials should be addressed to S.L. (email: s.lounis@fz-juelich.de).

\end{addendum}


\section*{References}
\bibliographystyle{naturemag}
\bibliography{references}

\begin{thebibliography}{10}
\expandafter\ifx\csname url\endcsname\relax
  \def\url#1{\texttt{#1}}\fi
\expandafter\ifx\csname urlprefix\endcsname\relax\def\urlprefix{URL }\fi
\providecommand{\bibinfo}[2]{#2}
\providecommand{\eprint}[2][]{\url{#2}}

\bibitem{Slater36}
\bibinfo{author}{Slater, J.~C.}
\newblock \bibinfo{title}{The ferromagnetism of nickel. ii. temperature
  effects}.
\newblock \emph{\bibinfo{journal}{Phys. Rev.}} \textbf{\bibinfo{volume}{49}},
  \bibinfo{pages}{931--937} (\bibinfo{year}{1936}).
\newblock \urlprefix\url{https://link.aps.org/doi/10.1103/PhysRev.49.931}.

\bibitem{Pauling38}
\bibinfo{author}{Pauling, L.}
\newblock \bibinfo{title}{The nature of the interatomic forces in metals}.
\newblock \emph{\bibinfo{journal}{Phys. Rev.}} \textbf{\bibinfo{volume}{54}},
  \bibinfo{pages}{899--904} (\bibinfo{year}{1938}).
\newblock \urlprefix\url{https://link.aps.org/doi/10.1103/PhysRev.54.899}.

\bibitem{Slater30}
\bibinfo{author}{Slater, J.~C.}
\newblock \bibinfo{title}{Cohesion in monovalent metals}.
\newblock \emph{\bibinfo{journal}{Phys. Rev.}} \textbf{\bibinfo{volume}{35}},
  \bibinfo{pages}{509--529} (\bibinfo{year}{1930}).
\newblock \urlprefix\url{https://link.aps.org/doi/10.1103/PhysRev.35.509}.

\bibitem{Slater30b}
\bibinfo{author}{Slater, J.~C.}
\newblock \bibinfo{title}{Atomic shielding constants}.
\newblock \emph{\bibinfo{journal}{Phys. Rev.}} \textbf{\bibinfo{volume}{36}},
  \bibinfo{pages}{57--64} (\bibinfo{year}{1930}).
\newblock \urlprefix\url{https://link.aps.org/doi/10.1103/PhysRev.36.57}.

\bibitem{BetheSommerfeld1933}
\bibinfo{author}{Sommerfeld, A.} \& \bibinfo{author}{Bethe, H.}
\newblock \emph{\bibinfo{title}{Elektronentheorie der Metalle}},
  \bibinfo{pages}{333--622} (\bibinfo{publisher}{Springer Berlin Heidelberg},
  \bibinfo{address}{Berlin, Heidelberg}, \bibinfo{year}{1933}).

\bibitem{Bogdanov1989}
\bibinfo{author}{Bogdanov, A.~N.} \& \bibinfo{author}{Yablonskii, D.}
\newblock \bibinfo{title}{Thermodynamically stable ``vortices" in magnetically
  ordered crystals. the mixed state of magnets}.
\newblock \emph{\bibinfo{journal}{J. Exp. Theor. Phys.}}
  \textbf{\bibinfo{volume}{95}}, \bibinfo{pages}{178} (\bibinfo{year}{1989}).

\bibitem{Roessler2006}
\bibinfo{author}{R\"ossler, U.~K.}, \bibinfo{author}{Bogdanov, A.~N.} \&
  \bibinfo{author}{Pfleiderer, C.}
\newblock \bibinfo{title}{Spontaneous skyrmion ground states in magnetic
  metals}.
\newblock \emph{\bibinfo{journal}{Nature}} \textbf{\bibinfo{volume}{442}},
  \bibinfo{pages}{797--801} (\bibinfo{year}{2006}).

\bibitem{Nagaosa2013}
\bibinfo{author}{Nagaosa, N.} \& \bibinfo{author}{Tokura, Y.}
\newblock \bibinfo{title}{{Topological properties and dynamics of magnetic
  skyrmions.}}
\newblock \emph{\bibinfo{journal}{Nat. Nanotech.}}
  \textbf{\bibinfo{volume}{8}}, \bibinfo{pages}{899--911}
  (\bibinfo{year}{2013}).

\bibitem{Fert2013}
\bibinfo{author}{Fert, A.}, \bibinfo{author}{Cros, V.} \&
  \bibinfo{author}{Sampaio, J.}
\newblock \bibinfo{title}{{Skyrmions on the track}}.
\newblock \emph{\bibinfo{journal}{Nat. Nanotech.}}
  \textbf{\bibinfo{volume}{8}}, \bibinfo{pages}{152--156}
  (\bibinfo{year}{2013}).

\bibitem{Fert2017}
\bibinfo{author}{Fert, A.}, \bibinfo{author}{Reyren, N.} \&
  \bibinfo{author}{Cros, V.}
\newblock \bibinfo{title}{{Magnetic skyrmions: advances in physics and
  potential applications}}.
\newblock \emph{\bibinfo{journal}{Nat. Rev. Mater.}}
  \textbf{\bibinfo{volume}{2}}, \bibinfo{pages}{17031} (\bibinfo{year}{2017}).

\bibitem{Shindou01}
\bibinfo{author}{Shindou, R.} \& \bibinfo{author}{Nagaosa, N.}
\newblock \bibinfo{title}{Orbital ferromagnetism and anomalous hall effect in
  antiferromagnets on the distorted fcc lattice}.
\newblock \emph{\bibinfo{journal}{Phys. Rev. Lett.}}
  \textbf{\bibinfo{volume}{87}}, \bibinfo{pages}{116801}
  (\bibinfo{year}{2001}).
\newblock
  \urlprefix\url{https://link.aps.org/doi/10.1103/PhysRevLett.87.116801}.

\bibitem{dosSantosDias2016}
\bibinfo{author}{dos Santos~Dias, M.}, \bibinfo{author}{Bouaziz, J.},
  \bibinfo{author}{Bouhassoune, M.}, \bibinfo{author}{Bl{\"u}gel, S.} \&
  \bibinfo{author}{Lounis, S.}
\newblock \bibinfo{title}{Chirality-driven orbital magnetic moments as a new
  probe for topological magnetic structures}.
\newblock \emph{\bibinfo{journal}{Nature Communications}}
  \textbf{\bibinfo{volume}{7}}, \bibinfo{pages}{13613} (\bibinfo{year}{2016}).

\bibitem{Hanke2016}
\bibinfo{author}{Hanke, J.-P.} \emph{et~al.}
\newblock \bibinfo{title}{Role of berry phase theory for describing orbital
  magnetism: From magnetic heterostructures to topological orbital
  ferromagnets}.
\newblock \emph{\bibinfo{journal}{Phys. Rev. B}} \textbf{\bibinfo{volume}{94}},
  \bibinfo{pages}{121114} (\bibinfo{year}{2016}).
\newblock \urlprefix\url{https://link.aps.org/doi/10.1103/PhysRevB.94.121114}.

\bibitem{dosSantosDias2017}
\bibinfo{author}{dos Santos~Dias, M.} \& \bibinfo{author}{Lounis, S.}
\newblock \bibinfo{title}{{Insights into the orbital magnetism of noncollinear
  magnetic systems}}.
\newblock \emph{\bibinfo{journal}{Spintronics X}}
  \textbf{\bibinfo{volume}{10357}}, \bibinfo{pages}{103572A}
  (\bibinfo{year}{2017}).
\newblock \urlprefix\url{https://doi.org/10.1117/12.2275305}.

\bibitem{Bouaziz2018}
\bibinfo{author}{Bouaziz, J.}, \bibinfo{author}{Dias, M. d.~S.},
  \bibinfo{author}{Guimar\~aes, F. S.~M.}, \bibinfo{author}{Bl\"ugel, S.} \&
  \bibinfo{author}{Lounis, S.}
\newblock \bibinfo{title}{Impurity-induced orbital magnetization in a rashba
  electron gas}.
\newblock \emph{\bibinfo{journal}{Phys. Rev. B}} \textbf{\bibinfo{volume}{98}},
  \bibinfo{pages}{125420} (\bibinfo{year}{2018}).
\newblock \urlprefix\url{https://link.aps.org/doi/10.1103/PhysRevB.98.125420}.

\bibitem{Lux2018}
\bibinfo{author}{Lux, F.~R.}, \bibinfo{author}{Freimuth, F.},
  \bibinfo{author}{Bl\"ugel, S.} \& \bibinfo{author}{Mokrousov, Y.}
\newblock \bibinfo{title}{Engineering chiral and topological orbital magnetism
  of domain walls and skyrmions}.
\newblock \emph{\bibinfo{journal}{Communication Physics}}
  \textbf{\bibinfo{volume}{1}}, \bibinfo{pages}{60} (\bibinfo{year}{2018}).

\bibitem{Brinker2020b}
\bibinfo{author}{Brinker, S.}, \bibinfo{author}{dos Santos~Dias, M.} \&
  \bibinfo{author}{Lounis, S.}
\newblock \bibinfo{title}{Spin, atomic, and interatomic orbital magnetism
  induced by $3d$ nanostructures deposited on transition metal surfaces}.
\newblock \emph{\bibinfo{journal}{Phys. Rev. Mater.}}
  \textbf{\bibinfo{volume}{4}}, \bibinfo{pages}{024404} (\bibinfo{year}{2020}).
\newblock
  \urlprefix\url{https://link.aps.org/doi/10.1103/PhysRevMaterials.4.024404}.

\bibitem{Tatara03}
\bibinfo{author}{Tatara, G.} \& \bibinfo{author}{Garcia, N.}
\newblock \bibinfo{title}{Quantum toys for quantum computing: Persistent
  currents controlled by the spin josephson effect}.
\newblock \emph{\bibinfo{journal}{Phys. Rev. Lett.}}
  \textbf{\bibinfo{volume}{91}}, \bibinfo{pages}{076806}
  (\bibinfo{year}{2003}).
\newblock
  \urlprefix\url{https://link.aps.org/doi/10.1103/PhysRevLett.91.076806}.

\bibitem{Bruno2004}
\bibinfo{author}{Bruno, P.}, \bibinfo{author}{Dugaev, V.~K.} \&
  \bibinfo{author}{Taillefumier, M.}
\newblock \bibinfo{title}{Topological hall effect and berry phase in magnetic
  nanostructures}.
\newblock \emph{\bibinfo{journal}{Phys. Rev. Lett.}}
  \textbf{\bibinfo{volume}{93}}, \bibinfo{pages}{096806}
  (\bibinfo{year}{2004}).
\newblock
  \urlprefix\url{https://link.aps.org/doi/10.1103/PhysRevLett.93.096806}.

\bibitem{Everschor14}
\bibinfo{author}{Everschor-Sitte, K.} \& \bibinfo{author}{Sitte, M.}
\newblock \bibinfo{title}{Real-space berry phases: Skyrmion soccer (invited)}.
\newblock \emph{\bibinfo{journal}{Journal of Applied Physics}}
  \textbf{\bibinfo{volume}{115}}, \bibinfo{pages}{172602}
  (\bibinfo{year}{2014}).
\newblock \urlprefix\url{https://doi.org/10.1063/1.4870695}.

\bibitem{Lux2020}
\bibinfo{author}{Lux, F.~R.}, \bibinfo{author}{Freimuth, F.},
  \bibinfo{author}{Bl\"ugel, S.} \& \bibinfo{author}{Mokrousov, Y.}
\newblock \bibinfo{title}{Chiral hall effect in noncollinear magnets from a
  cyclic cohomology approach}.
\newblock \emph{\bibinfo{journal}{Phys. Rev. Lett.}}
  \textbf{\bibinfo{volume}{124}}, \bibinfo{pages}{096602}
  (\bibinfo{year}{2020}).
\newblock
  \urlprefix\url{https://link.aps.org/doi/10.1103/PhysRevLett.124.096602}.

\bibitem{Bouaziz2021}
\bibinfo{author}{Bouaziz, J.}, \bibinfo{author}{Ishida, H.},
  \bibinfo{author}{Lounis, S.} \& \bibinfo{author}{Bl\"ugel, S.}
\newblock \bibinfo{title}{Transverse transport in two-dimensional relativistic
  systems with nontrivial spin textures}.
\newblock \emph{\bibinfo{journal}{Phys. Rev. Lett.}}
  \textbf{\bibinfo{volume}{126}}, \bibinfo{pages}{147203}
  (\bibinfo{year}{2021}).
\newblock
  \urlprefix\url{https://link.aps.org/doi/10.1103/PhysRevLett.126.147203}.

\bibitem{Nagaosa2019}
\bibinfo{author}{Nagaosa, N.}
\newblock \bibinfo{title}{Emergent inductor by spiral magnets}.
\newblock \emph{\bibinfo{journal}{Japanese Journal of Applied Physics}}
  \textbf{\bibinfo{volume}{58}}, \bibinfo{pages}{120909}
  (\bibinfo{year}{2019}).

\bibitem{Yokouchi2020}
\bibinfo{author}{Yokouchi, T.} \emph{et~al.}
\newblock \bibinfo{title}{Emergent electromagnetic induction in a helical-spin
  magnet}.
\newblock \emph{\bibinfo{journal}{Nature}} \textbf{\bibinfo{volume}{586}},
  \bibinfo{pages}{232} (\bibinfo{year}{2020}).

\bibitem{Dzyaloshinsky1958}
\bibinfo{author}{Dzyaloshinsky, I.}
\newblock \bibinfo{title}{A thermodynamic theory of “weak” ferromagnetism
  of antiferromagnetics}.
\newblock \emph{\bibinfo{journal}{J. Phys. Chem. Sol.}}
  \textbf{\bibinfo{volume}{4}}, \bibinfo{pages}{241 -- 255}
  (\bibinfo{year}{1958}).

\bibitem{Moriya1960}
\bibinfo{author}{Moriya, T.}
\newblock \bibinfo{title}{Anisotropic superexchange interaction and weak
  ferromagnetism}.
\newblock \emph{\bibinfo{journal}{Phys. Rev.}} \textbf{\bibinfo{volume}{120}},
  \bibinfo{pages}{91--98} (\bibinfo{year}{1960}).

\bibitem{Romming2013}
\bibinfo{author}{Romming, N.} \emph{et~al.}
\newblock \bibinfo{title}{Writing and deleting single magnetic skyrmions}.
\newblock \emph{\bibinfo{journal}{Science}} \textbf{\bibinfo{volume}{341}},
  \bibinfo{pages}{636--639} (\bibinfo{year}{2013}).

\bibitem{LimaFernandes2018}
\bibinfo{author}{{Lima Fernandes}, I.}, \bibinfo{author}{Bouaziz, J.},
  \bibinfo{author}{Bl{\"{u}}gel, S.} \& \bibinfo{author}{Lounis, S.}
\newblock \bibinfo{title}{{Universality of defect-skyrmion interaction
  profiles}}.
\newblock \emph{\bibinfo{journal}{Nat. Commun.}} \textbf{\bibinfo{volume}{9}},
  \bibinfo{pages}{4395} (\bibinfo{year}{2018}).

\bibitem{Fernandes2020}
\bibinfo{author}{Fernandes, I.~L.}, \bibinfo{author}{Bouhassoune, M.} \&
  \bibinfo{author}{Lounis, S.}
\newblock \bibinfo{title}{Defect-implantation for the all-electrical detection
  of non-collinear spin-textures}.
\newblock \emph{\bibinfo{journal}{Nature Commun.}}
  \textbf{\bibinfo{volume}{11}}, \bibinfo{pages}{1--9} (\bibinfo{year}{2020}).

\bibitem{Fernandes2020a}
\bibinfo{author}{Fernandes, I.~L.}, \bibinfo{author}{Chico, J.} \&
  \bibinfo{author}{Lounis, S.}
\newblock \bibinfo{title}{{Impurity-dependent gyrotropic motion, deflection and
  pinning of current-driven ultrasmall skyrmions in PdFe/Ir(111) surface}}.
\newblock \emph{\bibinfo{journal}{Journal of Physics: Condensed Matter}}
  (\bibinfo{year}{2020}).
\newblock \urlprefix\url{http://iopscience.iop.org/10.1088/1361-648X/ab9cf0}.

\bibitem{Fernandes22}
\bibinfo{author}{Fernandes, I.~L.}, \bibinfo{author}{Bl{\"u}gel, S.} \&
  \bibinfo{author}{Lounis, S.}
\newblock \bibinfo{title}{Spin-orbit enabled all-electrical readout of chiral
  spin-textures}.
\newblock \emph{\bibinfo{journal}{Nature Communications}}
  \textbf{\bibinfo{volume}{13}}, \bibinfo{pages}{1576} (\bibinfo{year}{2022}).
\newblock \urlprefix\url{https://doi.org/10.1038/s41467-022-29237-0}.

\bibitem{Lounis2007}
\bibinfo{author}{Lounis, S.}, \bibinfo{author}{Mavropoulos, P.},
  \bibinfo{author}{Zeller, R.}, \bibinfo{author}{Dederichs, P.~H.} \&
  \bibinfo{author}{Bl\"ugel, S.}
\newblock \bibinfo{title}{Noncollinear magnetism of cr and mn nanoclusters on
  ni(111): Changing the magnetic configuration atom by atom}.
\newblock \emph{\bibinfo{journal}{Phys. Rev. B}} \textbf{\bibinfo{volume}{75}},
  \bibinfo{pages}{174436} (\bibinfo{year}{2007}).
\newblock \urlprefix\url{https://link.aps.org/doi/10.1103/PhysRevB.75.174436}.

\bibitem{Bauer2013}
\bibinfo{author}{Bauer, D. S.~G.}
\newblock \bibinfo{title}{Development of a relativistic full-potential
  first-principles multiple scattering green function method applied to complex
  magnetic textures of nano structures at surfaces}.
\newblock \emph{\bibinfo{journal}{PhD dissertation at the RWTH-Aachen}}
  (\bibinfo{year}{2013}).

\bibitem{Crum2015}
\bibinfo{author}{Crum, D.~M.} \emph{et~al.}
\newblock \bibinfo{title}{{Perpendicular reading of single confined magnetic
  skyrmions}}.
\newblock \emph{\bibinfo{journal}{Nat. Commun.}} \textbf{\bibinfo{volume}{6}},
  \bibinfo{pages}{8541} (\bibinfo{year}{2015}).

\bibitem{Arjana2020}
\bibinfo{author}{Arjana, I.~G.}, \bibinfo{author}{Lima~Fernandes, I.},
  \bibinfo{author}{Chico, J.} \& \bibinfo{author}{Lounis, S.}
\newblock \bibinfo{title}{Sub-nanoscale atom-by-atom crafting of
  skyrmion-defect interaction profiles}.
\newblock \emph{\bibinfo{journal}{Scientific Reports}}
  \textbf{\bibinfo{volume}{10}}, \bibinfo{pages}{14655} (\bibinfo{year}{2020}).
\newblock \urlprefix\url{https://doi.org/10.1038/s41598-020-71232-2}.

\bibitem{Bouhassoune2021}
\bibinfo{author}{Bouhassoune, M.} \& \bibinfo{author}{Lounis, S.}
\newblock \bibinfo{title}{Friedel oscillations induced by magnetic skyrmions:
  From scattering properties to all-electrical detection}.
\newblock \emph{\bibinfo{journal}{Nanomaterials}} \textbf{\bibinfo{volume}{11}}
  (\bibinfo{year}{2021}).
\newblock \urlprefix\url{https://www.mdpi.com/2079-4991/11/1/194}.

\bibitem{Aldarawsheh22}
\bibinfo{author}{Aldarawsheh, A.} \emph{et~al.}
\newblock \bibinfo{title}{Emergence of zero-field non-synthetic single and
  interchained antiferromagnetic skyrmions in thin films}.
\newblock \emph{\bibinfo{journal}{Nature Communications}}
  \textbf{\bibinfo{volume}{13}}, \bibinfo{pages}{7369} (\bibinfo{year}{2022}).
\newblock \urlprefix\url{https://doi.org/10.1038/s41467-022-35102-x}.

\bibitem{Brinker2019}
\bibinfo{author}{Brinker, S.}, \bibinfo{author}{Dias, M. d.~S.} \&
  \bibinfo{author}{Lounis, S.}
\newblock \bibinfo{title}{The chiral biquadratic pair interaction}.
\newblock \emph{\bibinfo{journal}{New Journal of Physics}}
  \textbf{\bibinfo{volume}{21}}, \bibinfo{pages}{083015}
  (\bibinfo{year}{2019}).

\bibitem{Lounis2020}
\bibinfo{author}{Lounis, S.}
\newblock \bibinfo{title}{Multiple-scattering approach for multi-spin chiral
  magnetic interactions: application to the one- and two-dimensional rashba
  electron gas}.
\newblock \emph{\bibinfo{journal}{New Journal of Physics}}
  \textbf{\bibinfo{volume}{22}}, \bibinfo{pages}{103003}
  (\bibinfo{year}{2020}).
\newblock \urlprefix\url{https://dx.doi.org/10.1088/1367-2630/abb514}.

\bibitem{Grytsiuk2020}
\bibinfo{author}{Grytsiuk, S.} \emph{et~al.}
\newblock \bibinfo{title}{Topological chiral magnetic interactions driven by
  emergent orbital magnetism}.
\newblock \emph{\bibinfo{journal}{Nature Communications}}
  \textbf{\bibinfo{volume}{11}}, \bibinfo{pages}{1--7} (\bibinfo{year}{2020}).

\bibitem{Brinker2020}
\bibinfo{author}{Brinker, S.}, \bibinfo{author}{dos Santos~Dias, M.} \&
  \bibinfo{author}{Lounis, S.}
\newblock \bibinfo{title}{Prospecting chiral multisite interactions in
  prototypical magnetic systems}.
\newblock \emph{\bibinfo{journal}{Phys. Rev. Research}}
  \textbf{\bibinfo{volume}{2}}, \bibinfo{pages}{033240} (\bibinfo{year}{2020}).
\newblock
  \urlprefix\url{https://link.aps.org/doi/10.1103/PhysRevResearch.2.033240}.

\bibitem{Shehada22}
\bibinfo{author}{Shehada, S.}, \bibinfo{author}{dos Santos~Dias, M.},
  \bibinfo{author}{Abusaa, M.} \& \bibinfo{author}{Lounis, S.}
\newblock \bibinfo{title}{Interplay of magnetic states and hyperfine fields of
  iron dimers on mgo (001)}.
\newblock \emph{\bibinfo{journal}{Journal of Physics: Condensed Matter}}
  \textbf{\bibinfo{volume}{34}}, \bibinfo{pages}{385802}
  (\bibinfo{year}{2022}).
\newblock \urlprefix\url{https://dx.doi.org/10.1088/1361-648X/ac8135}.

\end{thebibliography}

\end{document}